\documentstyle{article}
\renewcommand{\Im}{\rm Im}
\renewcommand{\d}{\delta}

\newcommand{\g}{\gamma}
\newcommand{\m}{\mu}

\newcommand{\e}{\varepsilon}
\newcommand{\ar}{\longrightarrow}

\newcommand{\w}{\omega}
\newcommand{\la}{\lambda}
\newcommand{\s}{\sigma}

\renewcommand{\a}{\alpha}
\begin{document}
\title{Protection of information in quantum databases}
\author{Y.Ozhigov}
\maketitle

\ \ \ \ \ Address: \ Department of mathematics, Moscow state technological

\ \ \ \ \ University "Stankin", Vadkovsky per. 3a, 101472, Moscow, Russia

\ \ \ \ \ E-mail: \ y@oz.msk.ru

\section{Summary}

The conventional protection of information by cryptographical keys
makes no sense if a key can be quickly discovered by an unauthorized 
person. This way of penetration to the protected systems was made possible
by a quantum computers in view of results of P.Shor (\cite{Sh}) and
L.Grover (\cite{Gr}). This work presents the method of protection of an information
in a database from a spy even he knows all about its control system and has 
a quantum computer, whereas a database can not distinguish between 
operations of spy and legal user. 

Such a database with quantum
mechanical memory plays a role of probabilistic oracle for some
Boolean function $f$ : it returns the value $f(a)$ for a query $a$ of user 
in time $O(N^2 \log^3 n )$, after that restors its initial state also in this time,
where $N$ is cardinality of $Dom f$. The software of database is independent
of a function $f$. Classical state of such a database must contain
the list of pairs ($a,f(a)),\ a\in Dom\ f$, taken in some order. Quantum mechanical 
principles allow to mix all these lists with different orders and the same
amplitude in one quantum state called normal, and perform all user's 
operations extracting $f(a)$ only in states of such a sort. 
Now if somebody S tries to learn $f(b)$ for $b\neq a$ then this
action so ruins the normal state that the legal 
user with high probability 
will not obtain a pair of the form $a,\ldots$ and hence 
the presence of S will be 
detected.
It is proved that for a spy the probability to learn $f(b)$ asymptotically
( when $N\ar\infty$ ) does not acceed the probability of its exposure.

Here advantage is taken of relative diffusion transforms (RDT), which
make possible to fulfill all operations in normal states. Such transforms 
look like diffusion transforms applied by L.Grover in \cite{Gr} but 
RDT are defined in a quite different manner.

A classical database with property of such a sort is impossible.

\section{Introduction. The main definitions}

All known models of quantum computers: quantum Turing Machines (
look in \cite{De}, \cite{BV} ), quantum circuits (look in \cite{Ya}) and quantum
cellular automata (look in \cite{Wa})  can simulate each ather with a 
polynomial slowdown and have the same computational power as 
classical computers. It is unknown is it possible to simulate 
absolute (without oracles) quantum computations by a classical 
computer with a polynomial slowdown or not. Such a simulation is known
only with exponential slowdown (look in \cite{BV}). As for 
relativized (with oracles) computations, the classical simulation
with a polynomial slowdown is impossible (look in \cite{BB}) and 
there is much evidence that quantum computers are substantially 
more effective than any classical device (look in
\cite{DJ}, \cite{BB}, \cite{Sh},
\cite{Gr}).

 To create databases we shall use the model of quantum computer with two 
parts: classical part, which transforms by classical lows (say as Turing Machine or cellular
automaton), and quantum part which transforms by the quantum mechanical 
principles. We proceed with the exact definitions.

{\bf Memory} ({\it quantum part}). It is a set 
$\cal E$ which elements are called
qubits. $\cal E$ may be designed as a discrete lattice: $\cal E \subseteq\sc Z
 ^m$ or as a tree, etc. Each qubit takes values from the complex 1-dimensional sphere of
 radius 1: $\{ z_0 {\bf 0} +z_1 {\bf 1} \ | \ z_1 ,z_2 \in {\tt C}, |z_0 |^2+|z_1 |^2
 =1\}$. Here $\bf 0$ and $\bf 1$ are referred as basic states of qubit and 
form the basis
 of ${\tt C}^2$.
It will be convenient to divide $\cal E$ into registers of $2$ neighboring qubits each
so that each register takes values from $\w =\{ 0,1,2,3\}$.

A basic state of the quantum part is a function of the form
$e:\ {\cal E} \ar\{ 0,1\}$. If we fix some order on ${\cal E} 
=\{ \nu_1 ,\nu_2 ,\ldots , \nu_r \}$ ($r$ even), the basic state $e$ may be encoded as
$|e(\nu_1 ),e(\nu_2 ),\ldots ,e(\nu_r ) \rangle$. Such a state 
can be naturally identified with the corresponding word in alphabet $\w$.

Let $e_0 ,e_1 ,\ldots ,e_{K-1}$ be all basic states, taken in some fixed order,
$\cal H$ be $K$-dimensional Hilbert space with orthonormal basis $e_0 ,e_1 ,\ldots ,e_{K-1},\ 2^r =K$. This Hilbert space can be regarded as tensor product
${\cal H} _1 \bigotimes {\cal H}_2 \bigotimes\cdots\bigotimes{\cal H}_r$ of $2$-dimensional spaces, where ${\cal H}_i$ is generated by the possible values of $e(\nu _i ),\ \ {\cal H}_i \cong {\tt C}^2$. A (pure) state of quantum part is such element
$x\in \cal H$ that $|x|=1$. Thus, in contrast to classical devices, quantum device may be not only in basic states, but also in coherent states, and this imparts surprising properties to such devices.

Put ${\cal K} =\{ 0,1,\ldots ,K-1\}$. For elements $x=\sum\limits_{s\in\cal K} \la _s e_s ,\
y=\sum\limits_{s\in\cal K} \mu _s e_s \in \cal H$ their dot product $\sum\limits_{s
\in\cal K} \la _s \bar\mu _s$ is denoted by $\langle x|y\rangle $, where $\bar\m$ means complex conjugation of $\mu\in\tt C$, hence $\langle x|y\rangle =\overline{\langle y|x\rangle }$.

{\bf Unitary transformations}. Let $\{ 1,\ldots ,r\} =\bigcup_{i=1}^l L_i^s ,\ \ L_i^s \cap
L_j^s =\emptyset \ (i\neq j)$, unitary transform $U_i^s$ acts on $\bigotimes_{j\in L_i^s} e_j$,
then $U^s =\bigotimes_{i=1}^l U_i^s$ acts on ${\cal H} ,\ \ s=1,2,\ldots ,M$. 
We require that all $U_i^s
$ belong to some finite set of transformation independent of $\cal E$ which
can be easily performed by physical devices. 

{\it A computation} is a chain of such unitary transformations:

$$ \chi_0 \stackrel{U^1}{\ar}\chi_1 \stackrel{U^2}{\ar}\ldots \stackrel{U^M}{\ar}
\chi_M .
$$

The passages $U^s \ar U^{s+1} \ s=1,\ldots ,M$ and the value $M$
 are determined by the classical algorithm
which points the partition $\bigcup L_i$ and chooses the 
transformations $U_i^s$ 
sequentially for each $s$. This algorithm is performed by
the {\it classical part} of computer.

{\bf Observations}. Let $\chi =\sum\limits_{s\in\cal K} \la_s e_s$ be some fix state of
computer, often $\chi =\chi_M$. If $A\in\{ 0,1\}^k$ is the list of
possible values
for the first $k$ qubits, then we put
$ B_A =\{ i \ |\ \exists a_{k+1} ,a_{k+2} ,\ldots ,a_r \in\{ 0,1\}
 :\ \ e_i =A a_{k+1} a_{k+2} \ldots a_r \}$.  A (quantum) result of this observation
is a new state $\chi ^A =\sum\limits_{i\in B_A} \frac{\la_i}{\sqrt{p_a}} e_i$,
where $p_A =\sum\limits_{i\in B_A} \la_i^2$.
The {\it observation} of the first register in state $\chi$ is the procedure
which gives the pair: $<$ classical word $A$ , quantum state $\chi^A$ $>$
with probability $p_A$ for any possible $A\in\{ 0,1\}^k$.
To receive such words $A$ is the unique way for anybody to learn the result
of quantum computations.

\section{Diffusion transform}

In this section we recollect some notions and ideas from the works \cite{Gr}
 and \cite{BBHT}.

Every unitary transformation $U:\ {\cal H} \ar \cal H$ can be represented 
by it's matrix $U=(u_{ij} )$ where $u_{ij} =\langle U(e_j )|e_i \rangle $ so 
that for $x=\sum\limits_{p\in\cal K}
 \la _p e_p ,\ U(x) =\sum\limits_{p\in\cal K} \la '_p e_p$ we have $\bar\la ' =U\bar\la$,
where $\bar\la ,\bar\la'$ are columns with elements $\la_p ,\ \la'_p$ 
respectively. The following significant
{\it diffusion } transform $D$ (introduced in \cite{Gr}) is defined by 
 $D=-WRW^{-1}$,  
where $W=U_1 \bigotimes U_2 \bigotimes \ldots\bigotimes U_r$, each $U_i$ acts on 
${\cal H}_i$ and has the matrix
$$
J=\left(\begin{array}{cc} 1/\sqrt 2\ &1/\sqrt 2 \\
1/\sqrt 2 \ &-1/\sqrt 2
\end{array}\right) ,
$$
and $R$ is the phase invertion of $e_0$.
For every state
$x=\sum\limits_{p\in\cal K} \la _p e_p$ an average amplitude is taken as $x_{av} =\sum\limits_{p\in\cal K} \la _p /K$.

\newtheorem{TE}{Proposition }
\begin{TE} {\rm (Grover , \cite{Gr} ).}
For every state $x$
\begin{equation}
\langle  e_p |x\rangle -x_{av} =x_{av} -\langle e_p |D(x)\rangle  .
 \end{equation}
\end{TE}
This means that $D$ is the inversion about average.

 One step of Grover's
algorithm is unitary transform $G=DR_t$ where $R_t$ is phase rotation 
of some target state $e_t$. Proposition 1 implies that the amplitude of 
$e_t$ grows approximately on $1/\sqrt K $ 
as a result of application of $G$ to the state $x_0 =(e_0 +e_1 +\cdots +e_{K-1} )/\sqrt K$. 

The following two notes about this algorithm have been done in the work
\cite{BBHT}.

1. If we have the set $T$ of target states of cardinality $|T|=K/4$,
then one step of so modified transform $G$ makes all amplitudes of the states
$e\notin T$ equal to zero.

2. Let $W'$ be any unitary transform satisfying
$W'(e_0 )=\frac{1}{\sqrt{K}}\sum\limits_{i\in\cal K} e_i$.
Then Proposition remains true if $W'$ is taken instead of $W$ in the definition of $D$.

\section{Relative diffusion transform}
 
In this section we introduce the generalization of diffusion transform - 
relative diffusion transform (RDT) which is the key notion for the construction
of the control system for our database. 

In what follows for the set $C=\{ 0,1,\ldots ,N-1\} ,\ N<K$, $C'$ 
denotes $\{ e_i \ |\ 0\leq i\leq N-1\}$.

 Fix the notation:
$ \chi_C =\frac{1}{\sqrt{N}} \sum\limits_{i\in C} e_i$. Let $M$ be unitary transformation
 of the form $\cal H\ar\cal H$
such that $M(e_0 )=\chi_C$. Such transform $M$ is called $C$-mixing. 
We do not require that
subspace, spanned by $C'$ is $M$-invariant.

{\bf Definition}. RDT($B$) is the following transform: $D_C =-MRM^{-1}$ where $R$ is defined above, $M$ is $C$-mixing.

The following Lemma generalizes Proposition.

\newtheorem{LE}{Lemma}
\begin{LE} $D_C$ does not change an amplitude $\mu_s$ of $e_s$ if 
$s\notin C$
and makes it $\frac{2A}{N} -\mu_s$ if $s\in C$, where 
$A=\sum\limits_{s\in C} \mu_s ,\ \ N=|C|$.
\end{LE}

{\it Proof}

At first note that for every $s\in\cal K$ $M^{-1} (e_s )=
\sum\limits_{i\in\cal K} \a_s^i e_i$, $\a_s^0 =1/\sqrt{N}$ for any $s\in C$ and 
$\a_s^0 =0$ for other $s$. Really, $\a_s^0 =\langle M^{-1} (e_s )\ |\ e_0 \rangle =
\langle e_s \ |\ M(e_0 )\rangle =1/\sqrt{N}$
in view of unitarity $M$. Now for $x=\sum\limits_{s\in\cal K} \mu_s e_s$ we have the following equations:
$$ 
\begin{array}{l}
D_C (x) =-MR_0 (\sum\limits_{s\in\cal K} \mu_s \sum\limits_{i\in\cal K}
 \a_s^i e_i )\\
\qquad{} =-M(\sum\limits_{s\in\cal K} \mu_s
\sum\limits_{i\in\cal K} \a_s^i e_i -2\sum\limits_{s\in C} e_0/\sqrt{N}
)\\
\qquad{} =-\sum\limits_{s\in\cal K} \mu_s e_s 
+\frac{2}{\sqrt{N}} \sum\limits_{s\in C} \mu_s 
\frac{1}{\sqrt{N}}\sum\limits_{j\in C} e_j\\
\qquad{} =-\sum\limits_{s\notin C} \mu_s e_s +\sum\limits_{j\in C} 
e_j (\frac{2A}{N} -\mu_j ).\ \ \Box
\end{array}
$$
 
\newtheorem{CO}{Corollary}
\begin{CO}  Let $C\subset {\cal K},\ |C|=N,\ T\subseteq C,\ |T|=N/4$, 
and $M$ is $C$-mixing.  Then
$ -MR_0 M^{-1} R_T (\chi _C )=\chi_T$.
\end{CO}

It is readily seen that the applying of $D_C R_T$ for $|T|=N/4$ doubles 
amplitudes in states of the form $\chi_C$, whereas Grover's algorithm
increases them only on constant ($O(1/\sqrt{K} $). 
Note that generally speaking, RDT($C$) can not be 
realized on a quantum computer for arbitrary subset $C\subset\cal K$ (look at \cite{Oz} ).
But in the next section we shall show how RDT can be realized in our specific
case: for the databases.

\section{Realization of RDT on quantum computer.
Control system for the database}

Let $f$ be a function of the form: $\{ 0,1\}^n \ar \{ 0,1\}^n$.
A {\it presentation} of $f$ is a basic state of the form 

$a_0 ,f(a_0 ),
a_1 ,f(a_1 ) , \ldots ,a_{N-1} ,f(a_{N-1} ),\g_1 ,\g_2, \ldots\ \ $, where $a_0 ,a_1 ,\ldots ,a_{N-1}$ are
all different strings from $\{ 0,1\}^n$ taken in some order, $N=2^n$,
 
$\g =(\g_1, \g_2 ,\ldots , \g_{2nN} )$ are values of ancillary qubits.
There are $M=N!$ forms of presentations different only in ancillary qubits, we denote them by $P_0^\g ,P_1^\g ,\ldots ,P_{M-1}^\g$,
where $P_0$ corresponds to the lexicographic order on $\{ 0,1\}^n$. Notation:
$P_i =a_0^i ,f(a_0^i ),\ldots$, we shall omitt $\g =\bar 0$ in notations.
A string $B_a =(a,f(a))$ is called block. Put ${\cal M}=\{ 0,1,\ldots ,M-1\}$.

{\it Control system} of the database consists of the following two parts:

{\bf Preparation of the main state} This is a unitary transformation
$$
P_0 \ \ar\ \frac{1}{\sqrt{M}} \sum\limits_{i\in\cal M} P_i \stackrel{def}{=} \chi_0 .
$$

{\bf Extracting and restoring procedures} Given a query $a$ an extracting
procedure consists of two parts:

1. Unitary transform 
$$
Ex\ :\ \chi_0 \ar 
\chi_a \stackrel{def}{=}\frac{1}{\sqrt{ (N-1)!}}\sum\limits_{i\in\zeta (a)} P_i ,
$$
where $\zeta (a)=\{ i\ |\ a_0^i =a\}$.

2. Following observation of the first $2n$ qubits. This observation
gives the required information $a,f(a)$ with certainty and does not change 
the observed state because $\chi_a$ has the form $|a,f(a)\rangle\bigotimes\chi_a '$.

The restoring procedure is $(Ex)^{-1}$ which gives again $\chi_0$ and the database
is ready for the following query.

We shall describe only $Ex$ because the main state
can be prepared along similar lines. If $a=\s_1\s_2 \ldots \s_{n/2}$ 
is some query to the database, all $\s_i \in\w$,
then $C_j$ denotes the set of all basic states of the form $P_j$ where
 $a_0^j =\s_1\s_2\ldots \s_j \d_{j+1} \d_{j+2}\ldots \d_{n/2}$, all $\d_k \in\w$,
 $n$ even.
Given some RDT$(C_j )$: $D_j$ , the sequential application of $D_j R_{C_{j+1}}$ for $j=1,2,\ldots ,\frac{n}{2} -1$ results in $\chi_a$ in view of Corollary.
Now to complete the construction of $Ex$ it would suffice to realize some
$C_j$-mixing transform $M_j$ on a quantum computer. $M_j$ will be constructed in 
3 steps. Here we can not apply Walsh -Hadamard transform like in the work \cite{Gr} because $P_j$ do not exhauste all basic states of $\cal H$.

{\bf Step 1.} Given $e_0 =P_0 =B_0 ,B_1 ,\ldots ,B_{N-1} ,0,0,\ldots ,0$, at first we 
create the state $\xi =\ |\ B_0 ,\ldots ,B_{N-1} \rangle\bigotimes\chi_{H_0} \bigotimes\chi_{H_{N-1}} \bigotimes\chi_{H_{N-2}} \bigotimes\chi_{H_1}$,
where $H_l =\{ 0,1,\ldots , l-1\} $
if $l\neq 0$ and $H_0=\{ 0,1,\ldots ,2^{n-2j} \}$.
This can be done by independent applying to the ancillary registers the
transformations $|0\rangle \ar\chi_{H_l}$ built by A.Kitaev in the work \cite{Ki}.

Given a pair of sequences $\bar i =i_0 ,i_1 ,\ldots ,i_{N-1};
\ \ \bar r =r_0 ,r_1 ,\ldots ,r_{N-1},$ where $r_s \in H_{N-s} \ s=1,\ldots ,N-1,\ r_0
 \in H_0$, we define the pair of sequences: $k_0 ,k_1
 ,\ldots ,k_s ;\ \ h_{s+1}^s ,\ldots , h_{N-1}^s$ by induction on $s$, where $k_i ,\
 h_i^s$ depend on $\bar i$ and $\bar r$.

{\it Basis}. $s=0$. $k_0 =j_s ,\ \ h_1^0 ,\ldots ,h_{N-1}^0$ is obtained from $\bar i$ by deleting of $j_{r_0}$.

{\it Step}. $s>0$. All $k_0 ,\ldots , k_{s-1}$ are already defined. Put $k_s =h_{s+r_s}^{s-1}$, the new sequence $h_{s+1}^s ,\ldots ,h_{N-1}^s$ is obtained from $h_s^{s-1} ,\ldots ,h_{N-1}^{s-1}$ by deleting of $k_s$. 
Denote $0,1,\ldots ,N-1$ by $\bar 1$, $0,0,\ldots ,0$ by $\bar 0$.

Let $T=2^{n-2j} ,\ j_1 <j_2 <\ldots < j_T ,\ \ B_{j_1} ,b_{j_2} ,\ldots ,B_{j_T}$ be all blocks from $C_j$.

{\bf Step 2}. It is the chain of classical transformations (with unitary matrices
containing only ones and zeroes): $\xi\ar\xi_0\ar\ldots \xi_{N-1}$, where

$\xi _s =B_{k_0} ,B_{k_1} ,\ldots ,B_{k_{s-1}} ,B_{h_s^{s-1}} ,B_{h_{N-1}^{s-1}},\rho_0 ,\ldots ,\rho_{s-1} ,r_s ,\ldots ,r_{N-1}$.

{\bf A}. Passage $\xi\ar\xi_0$. It is the replacement of 
$r_0$ by the number $q$ such that $i_q =j_{r_0}$, where
$i_{r_0} =j_t$. This can be done in view of that 
the mapping $q\ar j_q$ is reversible.

{\bf B}. Passage $\xi_s \ar\xi_{s+1} ,\ \ s=0,1,\ldots ,N-2$. We find the
block $B_{k_s}$ and establish it immediately after $B_{k_{s-1}}$, 
the order of all other blocks remains unchanged. In view of the definition of $k_s$ this can be done by means of classical unitary transform
independently of the contents of blocks. Replacement $r_s \ar \rho_s$ ensure the reversibility, e.g. unitarity of this transformation.

{\bf Step 3} (Optional). Transform 
$\rho_s (\bar i ,\bar r )\ar\rho_s (\bar i ,\bar r ) 
-\rho_s (\bar 1 ,\bar 0 ),$ 

$s=0,1, \ldots ,N-1$ results in zeroes in ancillary qubits if an initial state is $P_0$. These steps been applied to $P_0$ give any states from $C_j$ with the same amplitudes, therefore they give $\chi_a$. 
$\Box$

More detailed analysis gives that steps 1-3 take the time $O(N^2 \log^2 N + T (N))$ on a quantum Turing Machine where $T (N)$ is the time required for the Step 1 when precision is fixed.
Hence the procedure $Ex$ takes the time 
$O((N^2 \log ^2 N +T (N))\log N )$.

We have described the procedure of extracting $a,f(a)$. The reverse prosedure 
restores the main state  of the database. The main state $\chi_0$ can be prepared 
along similar lines which takes $O(N^3 \log^3 N +NT(N)\log N)$ time.

Note that the observation of the first block described above gives $a,f(a)$ 
only in ideal case, e.g. if the following effects can be neglected.

1. Precision of transformations is not absolute, especially for the procedures in Step 1.
 
2. Presence of noice: spontaneous transformations of the forms: $0\ar 1,\ 0\ar-0$, which 
touch sufficiently small part of each block $B_i$.

3. Unauthorized actions. Some actions with the database
with the aim to learn a value $f(b)$ when the control system works at
the query $a\neq b$. We now turn to the point 3. In the last section
we shall briefly run through the point 2.

\section{Protection of information against unauthorized actions}

We presume that the aim of such actions is to learn $f(b)$ for $b\neq a$
with high probability $p$ and some $g$ blocs are inaccessible for these
 actions, the first block (where control system
observes the result) is among them.
 To do this would require to deal with $Np$
blocks of memory because values $f(b)$ are distributed among all blocks 
but the first
with the same 
probability at any instant of time. 

We shall regard the following scenario. Let somebody S (say, spy) be 
equipped with a quantum 
computer with its own memory. When our database is in state 
$\chi_{C_j}$ when
working on a query $a\ \ $ 
S fulfills the following:

a) observes any $Np$ accessible blocks of our database at one instant 
of time,
then

b) fulfills unitary transforms with the accessible part of the database and a memory
 his computer with the aim to cover up all traces of his observations.

After that the control system continues its work as usually. 
 Denote by 
 $P_{ex}$ the probability of that the control system 
 will not receive the word of the form 
$a,A$ when observing the first block (exposure of S).

\newtheorem{TH}{Theorem}
\begin{TH}
There exists a function $\a (g,N)$ such that $\forall \e >0\ \exists g:
\ \a (g,N)>1-\e\ \ N=1,2,\ldots $ with the following property.
For every choise of the block observed by S and his unitary
transformations
$$
P_{ex}  \geq p \a.
$$
\end{TH}

{\it Sketch of the proof}

The memory of computer used by S 
can be considered as ancillary part of memory in our database.

We shall write $\chi_i , \ R_i$ instead of $\chi_{C_i} ,\ R_{C_i}$.
Denote by $Q_0$ the state after unauthorized action with the state $\chi_j$. 
Then the control system
performs sequentially transformations $D_{i+j} R_{i+j+1} ,\ i=1,2,\ldots ,
t-j-1,\ t=n/2$, we denote by $Q_{i+1}$ its results: $Q_{i+1} =D_{i+j}
R_{i+j+1} (Q_i )$
and put $\e =\langle Q_0 \ |\ \chi_j \rangle $. In view of unitarity of 
all transformations at hand $\forall i= 1,\ldots ,t-j-1\ \ 
\langle Q_i \ |\ \chi_{i+j} \rangle =\e$.
Denote by $S_{suc}$ the set of such basic states, that the first block
has the form $a,A$ for some $A$. For any 
final state $Q_{t-j-1}$ the probability to expose S is $1-\sum\limits_{e\in S_{suc}}
|\langle e\ |\ Q_{t-j-1} \rangle |^2$. We have: 
$$
1-P_{ex} = p P_1 +(1-p)P_2 ,
$$ 
where
$P_1$ ($P_2$ ) is the probability that the
control system receives $a,A$ on condition that the block $a,f(a)$ was
observed by S, (was not observed by S respectively).
 
Case 1). The block $a,f(a)$ was observed by S

Let $L_i =(N-1)!2^{t-(i+j)}$ be the cardinality of $C_i$.
Denote by $q_i^{av}$ the average amplitude of all basic states from $C_{i+j}$
when database is in state $Q_i$. We thus have $|q_i^{av} |\leq |\e |/\sqrt{L_i}$.
 Let $\d_{norm}$ and $\d_{S}$ denote absolute growth
of average amplitudes among basic $C_t$ -states in cases
without S and with S respectively.  It follows from Lemma that $\d_{norm}\geq \e \d_{S}$
and in state $Q_{t-j-1}$ all basic states from $S_{suc}$ with nonzero amplitudes
contain in $C_t$. Therefore $P_1 \leq |\e |^2$.

Case 2). The block $a,f(a)$ was not observed by S.

Here we rouphly estimate $P_2 \leq 1$. Joining these cases we conclude that
$1-P_{ex} \geq p \e^2 +1-p$. At last, in view of assumed conditions 
$\e$ can be estimated as $ |\e |\leq 
2(1-p)^g$ . $\Box$
    
\section {Error correcting procedure for the database}

A random error in the database is a transformations on the basic states
induced by changes of qubits values of the forms $0\ar 1$ or vise versa and
changes of phases $0\ar -0$ or $1\ar -1$,
touching only small part of the qubits in each block of memory. Note that the 
phase errors : $0\ar -1 ,\ 1\ar -1$ can be reduced to the changes of values
as it is shown in the work \cite{CS} . Error correcting codes (ECC) is 
the conventional tool to correct errors of such a sort. Let each block contains
$n$ qubits. 

{\it Encoding} is an injection of the form : $E:\ \{ 0,1 \}^n \ar \{ 0,1\}^{n_1}$,
where $n_1 >n$. If $w_n (A) =\sum\limits_{i=1}^n a_i$ is Hamming weight of
the word $A=a_1 a_2 \ldots a_n \in \{ 0,1\}^n$, the distance between two such words
$A,B$ is $d_n (A,B) =w_n (A\bigoplus B)$ where $\bigoplus$ denotes a bitwise addition 
modulo 2. Put $d(E) =\min \limits_{A,B\in \{ 0,1\}^n} d_{n_1} (E(A) ,E(B))$. 
Then if 
for some $A' ,B' \in\{ 0,1\}^{n_1} \ B'\in\Im (E)\ \ d_{n_1} (A' ,B' )<d(E)/2$
then such $B'$ is defined for $A'$ uniquely and we obtain the partial functions
$A'\ar B'\ar E^{-1} (B')=A\in\{ 0,1\}^n$. Their superposition 
${\cal D} :\ A'\ar A$
is called decoding procedure for encoding $E$. $\cal D$ corrects $\leq d(E)/2$
errors occured in encoding words $B'$. This procedure is essentially classical
because the mapping $A'\ar B'$ is not reversible. But if we use additional registers
consisting of ancillary qubits and denote by $\g$ its contents we can regard a
reversible function $A' \ar B' ,\ \g (A' ) \ar E^{-1} (B' ) , \g (A' )$ 
instead of classical decoding and fulfill this procedure on quantum computer.
The work \cite{CS} also proposed simple and convenient quantum linear codes.
 
ECC can be used in course of computations to correct errors which occure
randomly as a result of noise. The size of ancillary register thus is the bigger
the time of computation is longer. The paper \cite{AB} presents error correcting
procedure which correct errors with constant rate repeatedly 
in course of computing and requires the size of ancillary registers polylogarithmical 
on the time of computing. This error correcting procedure can be applied to 
our database which results in basic states of the form $E(e_i ),\g_i$ instead of
$e_i$ considered below, here all properties of the database will remain unchanged.

\section{Acknowledgments}

I am grateful to Peter Hoyer for his comments and criticism and to Lov Grover
for his attention to my work.

\end{document}